# Flexoelectric control of a ferromagnetic metal

Wei Peng[1,2]*, Se Young Park[3]*, Chang Jae Roh[1,2,4], Junsik Mun[1,5], Hwiin Ju[4], Jinkwon Kim[1,2], Eun Kyo Ko[1,2], Zhengguo Liang[6], Sungsoo Hahn[1,2], Jinfeng Zhang[6], Liang Si[7], Yong Jin Jo[8], Tae Heon Kim[8], Changyoung Kim[1,2], Lingfei Wang[6], Miyoung Kim[1,5], Jong Seok Lee[4], Tae Won Noh[1,2]*, Daesu Lee[9,10]*

[1]Center for Correlated Electron Systems, Institute for Basic Science, Seoul 08826, Republic of Korea
[2]Department of Physics and Astronomy, Seoul National University, Seoul 08826, Republic of Korea
[3]Department of Physics and Origin of Matter and Evolution of Galaxies (OMEG) Institute, Soongsil University, Seoul 06978, Republic of Korea
[4]Department of Physics and Photon Science, Gwangju Institute of Science and Technology, Gwangju 61005, Republic of Korea
[5]Department of Materials Science and Engineering and Research Institute of Advanced Materials, Seoul National University, Seoul 08826, Republic of Korea
[6]Hefei National Laboratory for Physical Sciences at Microscale, University of Science and Technology of China, Hefei 230026, China
[7]Institut für Festkörperphysik, TU Wien, Vienna 1040, Austria
[8]Department of Physics and Energy Harvest-Storage Research Center (EHSRC), University of Ulsan, Ulsan 44610, Republic of Korea
[9]Department of Physics, Pohang University of Science and Technology, Pohang 37673, Republic of Korea
[10]Asia Pacific Center for Theoretical Physics, Pohang 37673, Republic of Korea

**Electric fields have played a key role in discovering and controlling exotic electronic states of condensed matter[1-3]. However, electric fields usually do not work in metals as free carriers tend to screen electrostatic fields. While a pseudo-electric field generated by inhomogeneous lattice strain, namely a flexoelectric field[4], can in principle work in all classes of materials, it remains experimentally unexplored in metals. Here, using heteroepitaxy and atomic-scale imaging, we show that flexoelectric fields can polarize a metallic oxide $SrRuO_3$ with unexpectedly large Ru off-center displacements. We also observe that the flexoelectrically induced polar state of $SrRuO_3$ leads to sizable lattice expansion, similar to the electrostrictive expansion caused by ionic displacements in dielectrics under an external electric field. We further suggest that flexoelectrically driven Ru off-centering promotes**

**strong coupling between lattice and electronic degrees of freedom, possibly enhancing the ferromagnetism of SrRuO$_3$. Beyond conventional electric fields, flexoelectric fields may universally engender novel electronic states and their control via pure atomic displacements in a nondestructive and fast manner.**

In-situ control of intrinsic material properties has been indispensable for condensed matter physics and materials science. Among the various approaches, the electrical means has shown great controllability and flexibility in manipulating a variety of physical properties such as phase transition[1], magnetism[2], and carrier dynamics[3]. To date, this approach has achieved significant success in the study of dielectrics and semiconductors. In a metal, however, abundant free charge carriers are supposed to screen electric fields substantially, thus rendering the conventional approaches of electric control impractical or largely inefficient, if any. Recent efforts have been devoted to electrolyte gating in the electric double-layer transistor via an electrostatic or electrochemical mechanism, but they usually suffer from various concerns such as limited penetration depth, reproducibility, reversibility, endurance, and speed[5].

Based on phenomenological analogies between strain gradients and electric fields, flexoelectric control of physical properties has recently emerged as a fascinating alternative to conventional electric control[6-10]. Flexoelectricity describes the generation of electric polarization $P_{\text{flexo}}$ in the presence of strain gradient $\partial u / \partial x$ as $P_{\text{flexo}} = \varepsilon f_{\text{eff}} (\partial u / \partial x)$, where $\varepsilon$ is the dielectric constant and $f_{\text{eff}}$ is the flexocoupling coefficient (Fig. 1a). Conceptually, this can be perceived as the polarization response of a medium to an applied flexoelectric field $E_{\text{flexo}} = f_{\text{eff}} (\partial u / \partial x)$. This flexoelectric field has been shown to resemble a conventional electric field in manipulating the properties of dielectrics and semiconductors, such as ferroelectricity[6,7], electrical and chemical states[8-10]. It is important to note that the flexoelectric field is not a real electric field that obeys Gauss's law but takes the essence of elastic fields. Thus, flexoelectric fields should be regarded as a pseudo-electric field, which may be free of electrostatic screening and serve as a unique means for controlling physical properties of a metal. However, despite recent theoretical predictions on a strong flexocoupling effect in metals[11,12], the existence of flexoelectricity in metals has yet to be experimentally validated.

Here we propose a mechanism to achieve large strain gradient in a metal, of which a flexoelectrically-induced polar structure is unambiguously observed. When a material with lower lattice symmetry is coherently grown on a substrate with higher lattice symmetry, the inherent structure distortion can give rise to a shear strain in the overlaid film. Provided that this shear strain gradually evolves at the nanoscale throughout the film (depending on the extent of interfacial coupling and lattice misfit), a giant strain gradient could occur without damaging the crystalline structure integrity. As will be discussed in detail later, this strain gradient flexoelectrically polarizes the bulk of $SrRuO_3$ (a ferromagnetic metal, henceforth SRO), leading to a so-called flexo-polar metal phase with substantial Ru off-center displacement.

**Theoretical examination of flexoelectricity in a metal**

We first constructed a supercell of cubic SRO in our theoretical simulation, as illustrated in Fig. 1b. The supercell consists of 18 unit-cells (u.c.) along the [100] axis (see Methods). A shear strain $u_{31}$ was imposed by artificially shearing Sr along the [010] axis while relaxing the rest ions. The shear strain propagates along the [100] direction with its magnitude varying in a sinusoid form (green curve in Fig. 1c). Accordingly, a shear strain gradient is derived from $\partial u_{31}/\partial x$ and plotted as the dark yellow curve in Fig. 1c. Surprisingly, we found sizable off-center displacements of both Ru and O, exceeding 3 and 6 pm under a strain gradient of $\sim 9\times10^6$ m$^{-1}$, respectively (Fig.

1d). The ionic off-centering corresponds to a unit-cell electrical dipole of 0.25 $e$ Å (assuming that Ru and O are purely ionic), which is comparable to those of other conventional ferroelectrics (e.g., ~1 $e$ Å in BaTiO$_3$). Notably, none of the SRO bulk structures under homogeneous lattice state exhibits polar instability in our phonon calculations, emphasizing the critical role of strain gradients to polarize SRO. Interestingly, we performed the same simulation on dielectric SrTiO$_3$ (henceforth STO), which yields similar dipole strength (Extended Data Fig. 1). This finding is consistent with a recent theoretical work of predicting comparable flexocoupling coefficients of a metal and a dielectric[11].

**Experimental design of strain gradients in SRO**

In its bulk form, SRO has a distorted, centrosymmetric perovskite structure with an oxygen octahedral tilting (OOT) pattern described by $a^-a^-c^+$ according to Glazer notation (where "–" represents antiphase tilting and "+" represents in-phase tilting). It forms an orthorhombic structure (*Pbnm*) below 820 K. Our first-principles calculations predicted that, under moderate compressive strain along the (111) orientation, the orthorhombic structure is energetically the most stable for SRO (Extended Data Fig. 2a). Besides, a monoclinic structure (*C2/c*) with a OOT pattern of $a^-a^-c^0$ and a rhombohedral structure ($R\bar{3}c$) with a OOT pattern of $a^-a^-a^-$ stay close in energy, possessing a small relative energy cost of 16.6 and 32.6 meV per formula unit, respectively.

To achieve large strain gradients in SRO, we consider a heterostructure of SRO epitaxially grown on a STO (111) substrate (Fig. 2a). Along the (111) orientation of perovskite, there exists strong interfacial coupling, including symmetry match and oxygen octahedral tilting, which exerts powerful control over the crystalline structure of the overlaid film[13,14]. Moreover, the coupling decreases gradually away from the interface, inducing certain structure relaxation. This provides a simple basis for producing a film with spatially evolving crystalline symmetry, which is naturally coupled with varying lattice strain. As illustrated in Extended Data Fig. 2b, on the (111) plane, the orthorhombic structure of SRO exhibits a large distortion of the regular hexagon network of A-site cations compared to that of a cubic structure; while the distortion substantially decreases for the monoclinic structure and is negligible for the rhombohedral structure. Therefore, it is highly likely that SRO adopts the rhombohedral structure at the interface of SRO/STO (111) as a consequence of interfacial coupling. Here we exclude the cubic structure of SRO in consideration of the (111) compressive strain. Within intermediate film thicknesses, SRO is expected to relax to monoclinic in response to weakened interfacial coupling. It can be also expected that the monoclinic SRO has three degenerate ferroelastic domains, exhibiting a macroscopic three-fold rotational symmetry around the out-of-plane direction following the three-fold rotational symmetry of substrate surface (Extended Data Fig. 2c).

Provided that the aforementioned structure relaxation occurs in a distance of several nanometers, a large shear strain gradient will be generated. This process is schematically illustrated in Fig. 2b. Owing to the inherent structural distortion, the main crystalline axis $[110]_{pc}$ (the subscript pc refers to pseudocubic, without further notice the subscript will be omitted below) tilts from the (111) plane by different angles in the varying structures, which is characterized by $\alpha$ in Fig. 2a. The values of $\alpha$ varies from 54.74° in cubic STO to 54.76° in rhombohedral SRO, and 56.14° in monoclinic SRO. When the SRO layer is coherently bonded on a cubic STO substrate, the contest between inherent structure distortion and the requirement of epitaxial coherence enforces a shear strain along the $[11\bar{2}]$ direction of the (111) plane. The shear strain $u_{31}$ can be calculated according to the schematic in Fig. 2c: $u_{31} = \delta x/z = \cot\alpha_1 - \cot\alpha_2$, where $\delta x$ is the cation shearing distance and $z$ is the (111) interplanar constant. $\alpha_1$ is the expected tilting angle of SRO that is presumed to be fully coherent with the atomic structure of STO. Therefore, $\alpha_1$ should take the value of STO, i.e. 54.74°. $α_2$ is the actual tilting angle of low-symmetry distorted SRO, namely 56.14° in the case of monoclinic structure. Therefore, the large $α$ difference between monoclinic SRO and cubic STO gives rise to a substantial shear strain of 3.5%. Assuming the structural relaxation within a thickness of 10 nm, a giant strain gradient of $3.5×10^6$ m$^{-1}$ can form.

To substantiate the proposed scenario, we grew high-quality SRO films on (111) STO substrates using pulsed laser deposition (PLD, see Methods). The film thickness was atomically controlled in the range of 8 to 65 unit cells (u.c., 1 u.c. is approximately 0.23 nm) via reflection high-energy electron diffraction (RHEED, Extended Data Fig. 3a and b). The high quality of the epitaxial heterostructure was verified by both X-ray diffraction (XRD) and atomic force microscopy (AFM, Extended Data Fig. 3c-e). In line with the expected monoclinic and rhombohedral structures, the SRO film possesses completely suppressed in-phase tilting (namely the $c^+$ component) according to half-integer Bragg diffraction by X-ray (Extended Data Fig. 4).

**Atomic-scale imaging of flexoelectricity in SRO**

We used scanning transmission electron microscopy (STEM) with the annular bright field (ABF) technique to directly visualize the atomic structure of SRO. Figure 2d displays an ABF-STEM image of the SRO/STO (111) heterostructure. The shearing of Sr cations in SRO away from the STO [110] axis (the red dotted line) is clearly observed, in line with our expectation. We extracted the Sr position along the [110] direction and plotted them in Extended Data Fig. 5, from which we evaluated the shear angle $α$ and the shear strain. Our calculation shows that the shear is small (~0.3%) at the interfacial region and gradually develops to a maximum of ~1.5% at the top. Therefore, the strain gradient across the film of ~ 6 nm is estimated to be ~$2×10^6$ m$^{-1}$, along the [11$\bar{2}$] direction.

Owing to such a large strain gradient, Ru in SRO is markedly displaced off-center along [110] (Fig. 2e), which is subject to a vector component of the strain gradient. This is consistent with our calculation, which shows the lowest energy cost along [110] (see Methods). Note that STO shows apparently a centrosymmetric structure. By mapping the displacement vectors of B-site cations (Fig. 2f), we revealed significant Ru displacement, exceeding 25 pm in the SRO film bulk. The corresponding unit-cell electrical dipole then exceeds 1 $e$ Å, on a par with that of $BaTiO_3$. Such a large electrical dipole warrants strong polarity of the SRO film. The large polarity from Ru off-centering is surprising since Ru contributes dominantly electrons at the Fermi level and should endure strong carrier screening of polar displacements. As a consequence, the flexo-polar metal phase of SRO may exhibit significant electron-polarity coupling. This is in sharp contrast to the previously reported Anderson-Blount polar metals such as $LiOsO_3$ and $NdNiO_3$ (ref. 14,15), of which the polarity is mainly generated from A-site cation displacements in a weak electron-phonon coupling scenario[16].

We further analyzed the OOT of SRO to investigate its local structures. As shown in Fig. 2e, the OOT of SRO manifests as a zigzag pattern of oxygen atoms, of which the zigzag angle $\theta$ can be extracted to quantify OOT. In Fig. 2g, we map $\theta$ of the same region by unit cell. The majority of the film away from the SRO/STO interface exhibits $\theta \sim 3°$. Noticeably, both OOT and polarity are significantly suppressed in the interfacial SRO region of ~5 u.c. (Fig. 2e), and then gradually saturated up to ~14 u.c. This finding supports the proposed scenario of interfacial coupling-induced structure evolution with thickness.

**Flexo-polar phase transition in SRO**

We used optical second harmonic generation (SHG) to investigate the structural symmetry of SRO in more details. Fig. 3a presents the SHG anisotropy patterns of an 11 u.c. SRO/STO (111) heterostructure using $P_{in}$-$P_{out}$ and $P_{in}$-$S_{out}$ configurations, as illustrated by the schematic (the data for other thicknesses can be found in Extended Data Fig. 6). The SHG patterns can be well simulated and fitted with the contribution of three electric dipoles in point group *m* (see Methods). The fitting is to reflect the existence of three symmetry-equivalent monoclinic domains that possess polar axes along [110], [101], and [011] in the STO coordinates, respectively. Furthermore, we plotted the SHG intensity $I^{PP}(2\omega)$, the peak intensity of the major lobe in the $P_{in}$-$P_{out}$ configuration, versus the thickness *t* in Fig. 3b. The plot exhibits a non-monotonic evolution: first increasing with thickness up to ~12 u.c, and then gradually decreasing. This trend suggests that the measured SHG is dominantly contributed by the film interior, consistent with the scenario of flexo-polarity. Therefore, these results confirm the emergence of a polar monoclinic phase in the bulk of SRO film. Moreover, the 5 u.c. SRO/STO yielded rather distorted SHG patterns with significantly suppressed intensity (Extended Data Fig. 6). This result again confirms the proposed rhombohedral structure at the interface, which should be lack of strain gradient and the associated flexo-polarity.

As the monoclinic structure distortion acts as a critical role of the flexo-polarity in SRO, it is conceivable that a polar-nonpolar phase transition would be coupled to a structure transition from monoclinic to a higher-symmetry structure at high temperatures. The higher-symmetry structure is likely to be centrosymmetric rhombohedral according to the calculated energy gain of different structures. To investigate this, we measured the temperature-dependent SHG intensity, $I^{PP}(2\omega)$, of the (111) SRO film (Fig. 3c). With temperature increasing, a transition seems to occur at ~630 K, which is followed by a temperature-independent plateau of weaker intensity. The finite value of the latter part can be explained by the electric quadrupole contribution of point group $\bar{3}m$, corresponding to the centrosymmetric rhombohedral phase (Extended Data Fig. 7).

Further with temperature-dependent XRD measurements, we show that this polar to nonpolar phase transition is coupled to a structure transition, as evidenced by a kink at ~ 630 K in the evolution of the (111) lattice constant of SRO (Fig. 3d). The non-linear evolution of lattice constant below 630 K is a consequence of the charge-lattice interaction, namely electrostrictive-like expansion caused by Ru off-centering along the polar axis. The evolution can be described by a thermodynamically derived equation[17]: $d_{(111)}(T) = d_{(111)}^{bulk}(T) \times [1 + aP(T)^2]$, as shown by the red solid line. In the equation, $d_{(111)}(T)$ is the lattice constant of the SRO film and $d_{(111)}^{bulk}(T)$ is the lattice constant of the non-polar bulk SRO, which linearly depends on temperature due to thermal expansion (the red dashed line). $P(T)$ is the polarity of the SRO film, of which its dependence on temperature can be extracted from the temperature-dependent SHG curve and $a$ is a constant (related to elastic constants). This finding corroborates our proposed scenario where the flexo-polarity of SRO is coupled to the low-symmetry structure distortion.

Moreover, the transition in the evolution of SHG intensity can be well described by a critical behavior: $I^{pp}(2\omega) \propto (T - T_f)^{\mu}$, where $T_f$ is the transition temperature and $\mu$ is a critical exponent (red line in Fig. 3c). The fitting yields $T_f$ ~ 630 K and $\mu$ ~ 1.4, in stark contrast to a critical exponent of 0.5 that is predicted for conventional ferroelectric phase transition according to the Landau theory[18]. The deviation is caused by the fact that the polarity does not act as the primary order parameter to drive the phase transition of SRO, but rather a consequence of the non-polar low-symmetry structure distortion as discussed.

**Flexoelectric control of ferromagnetism in SRO**

SRO is a well-recognized correlated electron system that exhibits strong coupling among lattice, charge, spin, and orbital degrees of freedom. Therefore, the markedly large Ru off-center displacement is expected to profoundly impact the electronic properties of SRO, which can be manifested in its magnetism[19,20]. Indeed, our first-principles calculations, along with transport and magnetization measurements, suggests a scenario wherein the effective electron correlation is boosted via Ru off-centering-driven electronic bandwidth reduction, so as to enhance ferromagnetism, such as saturated magnetic moment and ferromagnetic Curie temperature (see Methods, Extended Data Fig. 8 and 9). Furthermore, we find that the tremendous magnetostriction of SRO, which has been reported previously[21,22], probably cooperates with the boosted electron correlation to give rise to a strong coupling between magnetic anisotropy and the flexo-polarity.

Via magnetic field angle-dependent Hall effect measurements, we determined the magnetic easy axes (MEAs) of the polar SRO film to be along <110>, the same as the polar axes. By rotating the magnetic field $H$ in the $(11\bar{2})$ plane, the anomalous Hall resistance $R_{xy}$ displays two hysteretic jumps centered at $\theta_H = 58°$, 90°, and 122° with a period of 180°, respectively (Fig. 4a). These jumps result from the magnetization reversal, which should occur when a large magnetic field rotates over 90° from the initial magnetization direction. Hence, the MEAs are inferred to tilt by around -32°, 0°, and 32° from the [111] axis in the $(11\bar{2})$ plane, which happen to coincide with the three polar axes <110>. The MEA geometry is corroborated by a similar measurement with $H$ rotating in the $(1\bar{1}0)$ plane (not shown here). In addition, the three-fold symmetry of magnetization reversal with $H$ rotating in the (111) plane verifies the magnetic domain degeneracy, consistent with the structural domain symmetry (Fig. 4b). Moreover, we confirmed that the dominant MEAs remain consistent throughout a thickness range of 3.5 to 15 nm (all exhibit strong polarity) by the same set of measurements.

The magnetic anisotropy of SRO is known to be markedly impacted by a variety of factors, including strain, OOT, and defects[23-26]. On one hand, the bulk orthorhombic SRO is known to possess uniaxial magnetocrystalline anisotropy along the major crystalline axis [110] (Fig. 4c). On the other hand, (001) SRO films have been reported with a deviated MEA between the directions of [001] and [110] due to the various competing factors[24]. A similar situation would be expected in the (111) SRO film without flexo-polarity (Fig. 4d). Therein, the extraordinary coincidence of MEAs with the major crystalline direction <110> in polar SRO and their thickness-independence should imply a different factor that dominantly determines the magnetic anisotropy. To understand this result, we recall that an exceptional magnetostriction coefficient on the order of $10^{-3}$ has been previously identified in SRO (ref. 21,22). The magnetostriction effect describes the change of lattice shape under magnetization, and *vice versa*. Accordingly, the aforementioned electrostrictive-like strain due to Ru off-centering (~$5\times10^{-4}$ at 150 K, estimated from Fig. 3d) induces a gain of magnetic anisotropy energy, estimated to be ~1.3 meV per Ru (see Methods)[27]. This energy is much higher than the intrinsic magnetocrystalline anisotropy energy of SRO (ref. 28), which can thus facilitate the MEA to robustly align along the polar direction, i.e., <110>, as illustrated in Fig. 4e. In this process, electron correlation is expected to play an influential role as well[29]. While the detailed mechanism requires further extensive theoretical work to illustrate, the results point to an exciting potential of directly controlling magnetization direction via flexoelectricity.

**Outlook**

Our study demonstrates the spontaneous formation of strain gradients accompanying the inherent structure transition of SRO, which enables us to unambiguously observe the flexo-polar metal phase with a variety of controlled electronic properties. This work thus builds the foundation and promises many exciting opportunities for future fundamental research and applications of a pseudo-electric field, i.e., flexoelectric field, in metals. Especially, the previously well-established experimental schemes for applying flexoelectric fields have shown several outstanding characteristics including universality, non-destructiveness, and high speed. Thus, they can be adopted as a viable alternative to the electric control[4,30]. It can be envisioned that many electronic properties of metals can be flexibly controlled to create intriguing physics and functions such as strong magnetoelectric coupling[31], topological electronic states[32], and exotic spin textures[33].

**Methods**

**Theoretical calculation.** First-principles density functional theory calculations were performed within the local density approximation plus $U$ (LDA+$U$) scheme using the Vienna ab initio simulation package (VASP)[34,35]. The projector augmented wave (PAW) method was used with the Ceperley-Alder exchange correlation functional[36,37]. The on-site Coulomb interaction considered by rotationally invariant LDA+$U$ was parameterized by $U = 3$ eV and $J = 0.75$ eV for Ru-$d$ orbitals[38], which reproduces the experimental magnetic anisotropy of bulk SRO sensitive to the choice of the $U$ value. The choice of the parameters is consistent with other literatures[39,40]. We used the charge-only LDA exchange-correlation functional with plus $U$ extension, giving increasing exchange splitting proportional to $J$ for SRO[41] and other transition metal oxides[42,43]. We used the energy cutoff of 500 eV and $k$-point sampling of $6 \times 6 \times 4$ for the 20-atom $\sqrt{2} \times \sqrt{2} \times 2$ unit cell, $6 \times 6 \times 3$ and $3 \times 3 \times 3$ for the 30-atom $\sqrt{2} \times \sqrt{2} \times 2\sqrt{3}$ and the 120-atom $2\sqrt{2} \times 2\sqrt{2} \times 2\sqrt{3}$ supercells used in the strained-bulk calculations, respectively, and $1 \times 8 \times 8$ for 126-atom $18 \times 1 \times 1$ slab used to evaluate flexoelectric distortions. The atomic positions were relaxed with the force threshold of 0.02 eV/Å.

The flexoelectric distortions were calculated in a slab geometry with a vacuum of 25Å. Due to the calculation cost, the $18 \times 1 \times 1$ supercell was constructed based on 5-atom-cubic SRO with volume optimized with LDA+$U$ scheme, we found no ferroelectric instability in cubic structure. The initial atomic structure of the supercell was prepared with the $z$-position of the $i^{th}$ atom displaced in the $z$-direction by $\Delta z_i = h \times Na \sin\left(\frac{2\pi * x_i}{Na}\right)$, where $N$ is the number of unit-cell in the $x$-direction ($N=18$), $a$ is the cubic lattice constant (3.884 Å), $x_i$ is the position in the $x$-direction, and $h$ is the parameter controlling the sheer strain gradient[44]. The parameter $h$ was set to 0.03 to have maximum strain gradient about $9 \times 10^6$ m$^{-1}$, comparable with the experimental value ($3.5 \times 10^6$ m$^{-1}$). While fixing the Sr atoms at the initial position, the atomic positions of Ru and O atoms were relaxed, in which the off-centering displacements of Ru and O atoms from the initial position were measured. The flexoelectric distortion of STO was calculated in the same way with the cubic lattice constant of 3.860 Å.

To investigate the preferred direction for Ru off-centering, we calculated the energy cost of shifting Ru atoms by 0.05 Å to different pseudo cubic directions: [100], [010], [001], [110], [101], [011], [111], and [1$\bar{1}\bar{1}$]. The calculated energy cost per formula unit for these directions relative to the [110] direction is: 0.61, 0.61, 2.46, 0, 1.37, 1.37, 0.71, and 1.29 meV, respectively. Therefore, under the experimentally observed shear strain gradient, Ru is supposed to displace along the [110] direction, as experimentally observed.

The change in the partial density of states (PDOS) as a function of Ru displacement was evaluated by gradually shifting the Ru positions along the pseudo cubic [110] direction from those of relaxed *Imma* ($a^-a^-c^0$) structure. We calculated the non-magnetic PDOS of Ru-$d$ orbitals, which contribute dominantly the electronic states at the Fermi level and are mainly responsible for the electronic properties of SrRuO$_3$. We found that, evolving from the *Pbnm* phase to the *Imma* phase, the suppressed c$^+$ rotation induces bandwidth increasing and band splitting reduction, which can explain the generally observed ferromagnetism suppression in compressively strained SRO films[45,46]. By imposing Ru displacement with growing magnitudes, SRO shows continuously decreased bandwidth and amplified band splitting, suggesting enhanced effective electron correlation. Furthermore, the DOS peak near the Fermi level rises with its position skewed to the upper band edge. These characteristics can well explain the enhanced ferromagnetism of the flexo-polar phase of SRO[19,20].

**Sample fabrication and characterization.** SRO films were grown on (001) or (111) STO substrates using a pulsed laser deposition (PLD) system with a KrF excimer laser (248 nm). A reflection high-pressure high-energy electron diffraction (RHEED) system was used to monitor the growth. Before deposition, STO substrates (miscut < 0.1°) were etched with a buffered hydrofluoric acid solution and then annealed in air at 1050 °C for 1 h to produce atomically flat surface with unit-cell step terrace structure. The obtained substrates were further leached in DI water for 1 h to remove excess Sr on the surface. During deposition, the temperature of the substrate was maintained at 675 °C. The SRO ultrathin films were grown under an oxygen pressure of 100 mTorr with a laser fluence of 2 J cm$^{-2}$.

AFM was performed using a Cypher scanning probe microscope (MFP-3D, Asylum Research) with Ir-Pt-coated tips (PPP-EFM, Nanosensors). Reciprocal space mapping by XRD was performed with a Bruker D8 Discover. Half-integer Bragg diffraction and temperature-dependent ω-2θ scan was carried out using synchrotron XRD at the 3A beamline of Pohang Accelerator Laboratory. The (111) lattice constants of SRO and STO were extracted from Gaussian fitting of the (222) peaks.

**SHG measurement.** SHG measurements were performed with two femtosecond wave sources possessing a central wavelength of 800 and 840 nm, and a repetition rate of 80 MHz and 250 kHz, respectively. The polarization of fundamental wave and second harmonic wave was controlled by half wave plate and Glan-Taylor polarizer, respectively. We isolated SHG wave from fundamental wave by adopting low pass- and band pass-filters, and the isolated SHG wave is detected by a photomultiplier tube. Temperature-dependent measurements were carried out in an incident-plane-rotating setup with a high temperature stage (HCP621G; Instec)[47].

The SHG results were analyzed with both electric quadrupole (EQ) and electric dipole (ED) contributions, which are given by $I_{EQ}^{2\omega} \propto |E_{EQ}^{2\omega}|^2 = |\chi_{ijkl}E_j^\omega \partial_k E_l^\omega|^2$ and $I_{ED}^{2\omega} \propto |E_{ED}^{2\omega}|^2 = |\chi_{ijk}E_j^\omega E_k^\omega|^2$, respectively. Here, $E_i^\omega$ denotes electric field component of the fundamental wave with optical polarization along the $i$-axis; $\chi_{ijkl}$ and $\chi_{ijk}$ stand for second- and third- order susceptibility tensor components, respectively. For the room-temperature polar phase of SRO, we used the ED contribution of three equivalent polar monoclinic domains with a point group $m$ for fitting in consideration of the observed polar axis. For the high-temperature non-polar phase, we considered the EQ contribution of a non-polar rhombohedral structure with a point group $\bar{3}m$ for fitting.

In the following, we provide the allowed nonlinear susceptibility tensor components for the considered crystal symmetry and the details of analysis.

For the ED contribution of three equivalent polar monoclinic domains (point group $m$), The second-order susceptibility tensors for the monoclinic point group $m$ are:

$$\chi_{xxx}, \quad \chi_{xyy}, \quad \chi_{xzz}, \quad \chi_{xzx}, \quad \chi_{xxz}, \quad \chi_{yyz}, \quad \chi_{yzy},$$

$$\chi_{yxy}, \quad \chi_{yyx}, \quad \chi_{zxx}, \quad \chi_{zyy}, \quad \chi_{zzz}, \quad \chi_{zzx}, \quad \chi_{zxz}$$

To consider the SHG wave in the sample coordinates (x', y', z'), we decompose the tangential components of the fundamental wave in the lab coordinates (x, y, z). The two coordinate systems can be transformed into each other according to the the conventional rule of three-dimensional coordinate rotation that is described by two angles α and β, where (x, y, z) is first rotated around the y axis by α and then around the z axis by β to get (x', y', z'). Therefore, the electric field components of fundamental wave inside the thin film ($E_{x'}^{\omega}$, $E_{y'}^{\omega}$, $E_{z'}^{\omega}$) are:

$$E_{x'}^{\omega} = (E_x^{\omega} \cos\alpha + E_z^{\omega} \sin\alpha) \cos\beta - E_y^{\omega} \sin\beta$$

$$E_{y'}^{\omega} = (E_x^{\omega} \cos\alpha + E_z^{\omega} \sin\alpha) \sin\beta + E_y^{\omega} \sin\beta$$

$$E_{z'}^{\omega} = -E_x^{\omega} \sin\alpha + E_z^{\omega} \cos\alpha$$

Then the electric field components of SHG wave contributed by ED ($E_{x'}^{2\omega}$, $E_{y'}^{2\omega}$, $E_{z'}^{2\omega}$) are:

$$E_x^{2\omega} = (E_{x'}^{2\omega} \cos\beta + E_{y'}^{2\omega} \sin\beta) \cos\alpha - E_{z'}^{2\omega} \sin\alpha$$

$$E_y^{2\omega} = E_{y'}^{2\omega} \cos\beta - E_{x'}^{2\omega} \sin\beta$$

$$E_z^{2\omega} = (E_{x'}^{2\omega} \cos\beta + E_{y'}^{2\omega} \sin\beta) \sin\alpha + E_{z'}^{2\omega} \cos\alpha$$

The SHG of three-equivalent monoclinic domains, of which their orientations in the laboratory coordinate are rotated by 120° around the z axis from each other, can be then obtained by coherently adding up the contribution of each domain: $I_{total}^{2\omega} = |E_{\theta}^{2\omega} + E_{\theta+120°}^{2\omega} + E_{\theta+240°}^{2\omega}|^2$.

For the EQ contribution of a centrosymmetric rhombohedral structure (point group $\bar{3}m$), the third-order susceptibility tensors include:

$$\chi_{zzzz}, \quad \chi_{xxxx} = \chi_{yyyy} = \chi_{xxyy} + \chi_{xyyx} + \chi_{xyxy},$$

$$\chi_{xxyy} = \chi_{yyxx}, \quad \chi_{xyyx} = \chi_{yxxy}, \quad \chi_{xyxy} = \chi_{yxyx},$$

$$\chi_{yyzz} = \chi_{xxzz}, \quad \chi_{zzyy} = \chi_{zzxx}, \quad \chi_{zyyz} = \chi_{zxxz}, \quad \chi_{yzzy} = \chi_{xzzx},$$

$$\chi_{xxxz} = -\chi_{xyyz} = -\chi_{yxyz} = -\chi_{yyxz},$$

$$\chi_{xxzx} = -\chi_{xyzy} = -\chi_{yxzy} = -\chi_{yyzx},$$

$$\chi_{xzxx} = -\chi_{xzyy} = -\chi_{yzxy} = -\chi_{yzyx},$$

$$\chi_{zxxx} = -\chi_{zxyy} = -\chi_{zyxy} = -\chi_{zyyx},$$

$$\chi_{zyzy} = \chi_{xzxz}, \quad \chi_{zyzy} = \chi_{zxzx}$$

The analytical expressions for SHG are shown below and used for fitting:

$$I^{PP}(2\omega) = A_1\left[\left(\chi_{zyzy} - 2\chi_{zzyy} + \chi_{zzzz} + \chi_{zyyy}\sin(3\phi)\right)^2 + \left(2\chi_{xxyy} + \chi_{xyxy} - 2\chi_{xxxy} + \chi_{yzyz} + (\chi_{yyzy} - \chi_{yyyz})\sin(3\phi)\right)^2\right]$$

$$I^{PS}(2\omega) = A_2\left(\chi_{xyxy} - \chi_{yyzy}\sin(3\phi)\right)^2 + \left(\chi_{zyzy} + \chi_{zyyy}\sin(3\phi)\right)^2$$

$$I^{SP}(2\omega) = A_3\left((\chi_{yyzy} - 2\chi_{yyyz})\cos(3\phi)\right)^2$$

$$I^{SS}(2\omega) = A_4\left(\chi_{yyzy}\cos(3\phi)\right)^2$$

**STEM measurement and analyses.** The cross-sectional STEM specimen was prepared first by thinning the sample using a focused ion beam milling workstation (Helios NanoLab 650; FEI Co., Hillsboro, OR, USA) with low-energy ion beams (< 2 kV), then followed by focused low-energy (< 500 eV) Ar-ion milling (NanoMill 1040; E.A. Fischione Instruments, Export, PA, USA). Zone-axis $[1\bar{1}0]$ was predetermined by rotational in-plane XRD.

Atomic-resolution ABF-STEM experiments were performed at room temperature using spherical aberration probe-corrected STEM with an acceleration voltage of 200 kV (JEM-ARM 200F; JEOL Ltd., Tokyo, Japan). The instrument was equipped with a cold-field emission gun installed at the National Center for Inter-University Research Facilities (NCIRF), Seoul National University, South Korea. The TEM specimen was cleaned using ion cleaner (JEC-4000DS; JEOL Ltd.) prior to the STEM experiments. For STEM analyses, the semi-convergence angle was set to 24 mrad and the semi-collection angle range was set to 12-24 mrad.

To minimize scan distortion and enhance the signal-to-noise ratio, 20 frames of serial STEM images were acquired, with a short dwell time of 2 μs/px. Each image was 1024 × 1024 in size. The image series was registered using both rigid and non-rigid methods, which yielded similar results[48,49]. The atomic positions were extracted using a 2D Gaussian fitting method with seven parameters, based on a customized MATLAB script. To eliminate the artifact induced by the tail of heavy elements, atomic positions were calculated by atomic mass order after removing the larger atomic peaks.

**Low-temperature magnetism and transport measurements.** Conventional photolithography and ion-milling were used to pattern the SRO films into the Hall bar geometry. The channel size was minimized to $50 \times 50$ μm$^2$. Pt (50 nm) electrodes were sputtered onto the Hall bar contacts to reduce contact resistance. Magnetization measurements were recorded using a SQUID magnetometer (MPMS; Quantum Design). The longitudinal and transverse resistance were measured using a physical properties measurement system (PPMS, Quantum Design) on standard Hall bars. Mangetic field angle-dependent Hall effect measurement was carried out with a home-made rotational stage.

For analyzing the ordinary Hall resistance, the high-field transverse resistivity of the polar (111) SRO is fitted using the two-carrier model: $\rho_{xy} = \frac{1}{e} \frac{(n_h \mu_h^2 - n_e \mu_e^2) + \mu_h^2 \mu_e^2 H^2 (n_h - n_e)}{(n_h \mu_h + n_e \mu_e)^2 + \mu_h^2 \mu_e^2 H^2 (n_h - n_e)^2} H$, where $n_h$, $\mu_h$, $n_e$, $\mu_e$ are the carrier density and mobility of hole and electron carriers, respectively; $H$ is the magnetic field; $e$ is the elementary charge. The fitting yields the ordinary Hall component (OHE, blue line), which can be then subtracted from the raw result to get the anomalous Hall component (AHE, red line), i.e. $\rho_{xy} = \rho_{OHE} + \rho_{AHE}$. For the analysis of the transverse resistivity of (001) SRO film, the OHE can be simply obtained from linear fitting of the high-field Hall resistivity, indicating its pure electron-type carrier. The fitting parameters are provided in Extended Data Fig. 8c. The fitting results show that the polar SRO possesses comparable electron density as that of the non-polar SRO but smaller electron mobility. Moreover, the polar SRO exhibits an additional high-mobility, low-density hole type carrier, which may suggest the impact of Ru displacements on the electronic bands.

For analyzing the magnetization of the polar SRO under a small in-plane and out-of-plane magnetic field (1000 oe), we assume that the three degenerated magnetic domains are equal in volume. By applying an out-of-plane magnetic field (Extended Data Fig. 9a), the magnetic domains are all poled tilted upward. Therefore, the out-of-plane magnetization $M_{OOP}$ will be $M_{OOP} = \frac{1}{3} M_s \cos 35° \times 3 = 0.819 M_s$, where $M_s$ is the saturated magnetization along the easy axis. By applying the magnetic field along different in-plane directions (Extended Data Fig. 9b), the magnetic domains can be poled differently as illustrated in the lower panel. In the case of H//[11$\bar{2}$], the in-plane magnetization component $M_{IP} = \frac{1}{3} M_s \sin 35° \cos 60° \times 2 + \frac{1}{3} M_s \sin 35° = 0.382 M_s$; in the case of H//[1$\bar{1}$0], $M_{IP} = \frac{1}{3} M_s \sin 35° \cos 30° \times 2 = 0.331 M_s$. As a result, the ideal ratio of $M_{IP}/M_{OOP}$ is 0.47 and 0.41 for the two cases, respectively. Note that variation of the actual sample orientation in the measurement will cause slight deviation from the ideal ratio.

To estimate the magnetic anisotropy energy gain from the magnetostriction effect, we used the phenomenological equation that was introduced for SRO (001) films in ref. 27: $\Delta E = 3\lambda(c_{11} - c_{12})(1 + \frac{2c_{11}}{c_{12}})\epsilon$, where $\lambda$ is the magnetostriction coefficient, $c$ is the elastic moduli, and $\epsilon$ is the strain. We assume that in our geometry, $\lambda$ is in the order of 10$^{-3}$; the elastic moduli and their difference in the order of 100 GPa; $\epsilon$ is the electrostrictive-like strain due to Ru off-centering, which is ~5×10$^{-4}$ at 150 K. Accordingly, $\Delta E$ is estimated to ~1.3 meV.

[47] Kim, S. H. *et al.* A compact and stable incidence-plane-rotating second harmonics detector. *Rev. Sci. Instrum.* **92**, 043905 (2021).
[48] Jones, L. *et al.* Smart Align—a new tool for robust non-rigid registration of scanning microscope data. *Adv. Struct. Chem. Imaging* **1**, 8 (2015).
[49] Savitzky, B. H. *et al.* Image registration of low signal-to-noise cryo-STEM data. *Ultramicroscopy* **191**, 56-65 (2018).

## Acknowledgements

This work was supported by the Research Center Program of IBS in Korea (IBS-R009-D1) and by the National Research Foundation of Korea(NRF) grant funded by the Korea government(MSIT) (No. 2021R1A5A103299611). STEM measurement was supported the National Center for Inter-University Research Facilities at Seoul National University. This work was supported by Samsung Electronics Co., Ltd (No. IO201211–08061-01). S.Y.P. was supported by the National Research Foundation of Korea (NRF) grant funded by the Korea government(MSIT) (No. 2021R1C1C1009494) and by Basic Science Research Program through the National Research Foundation of Korea (NRF) funded by the Ministry of Education (No. 2021R1A6A1A03043957). Y.J.J. and T.H.K. acknowledge support from the Priority Research Centers Program through the National Research Foundation of Korea (NRF) funded by the Ministry of Education (Grant No. NRF‐2019R1A6A1A11053838). Experiments at PLS-II were supported in part by MSIT and POSTECH. L.W. is supported by the National Basic Research Program of China (Grant Nos. 2020YFA0309100), the National Natural Science Foundation of China (Grant Nos. 12074365), and the Fundamental Research Funds for the Central Universities (Grant Nos. WK2340000102, WK2030000035),

## Author contributions

W.P., D.L., and T.W.N. conceived the ideal and designed the experiments. W.P. grew the materials, fabricated the devices, performed lab XRD, AFM, and magnetotransport measurements with the help from E.K.K. S.Y.P. performed first-principles calculations. C.J.R., H.J., and J.S.L. performed SHG measurements. J.M. and M.K. performed STEM measurements. W.P., J.K., E.K.K., Y.J.J., and T.H.K. performed synchrotron XRD. S.H., S.L., and C.K. performed band structure analysis. W.P., Z.L., J.Z., and L.W. performed magnetization measurement. W.P. and D.L. analyzed the data and wrote the manuscript, with input from all authors.

## Corresponding authors

wei.peng@snu.ac.kr
sp2829@ssu.ac.kr
twnoh@snu.ac.kr
dlee1@postech.ac.kr

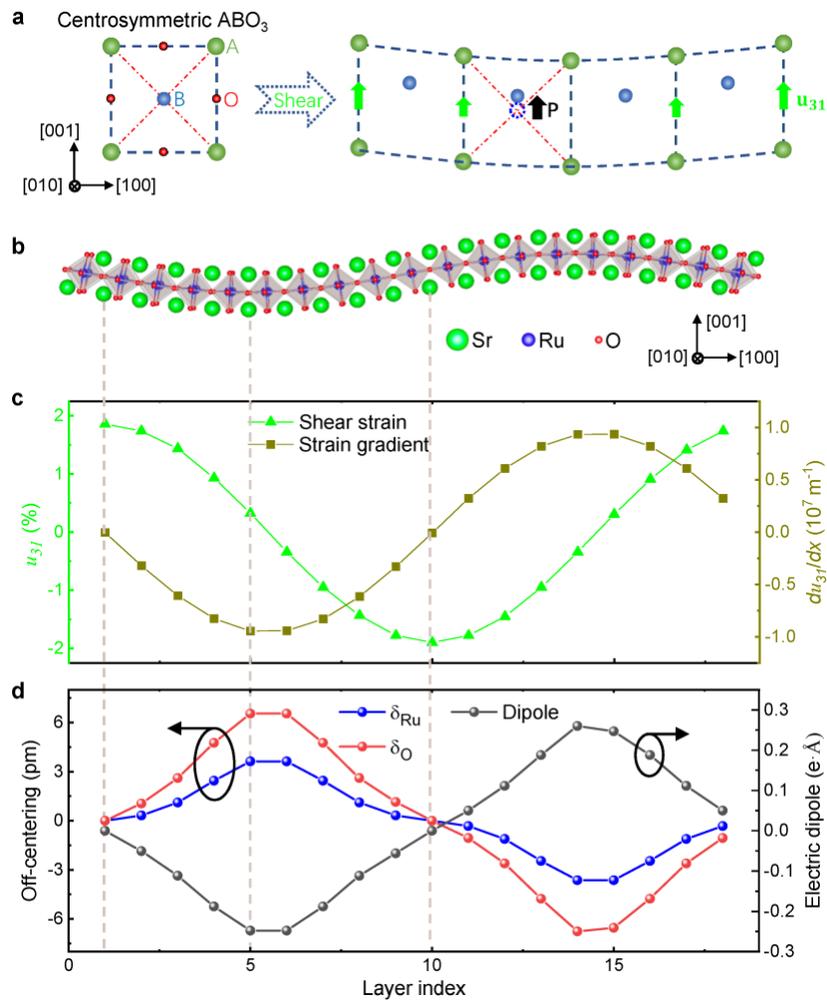

**Figure 1. First-principles calculation showing the existence of flexoelectricity in SRO. (a)** Schematic to show the flexoelectricity that is generated from a shear strain gradient in perovskite oxides ABO$_3$. **(b)** The supercell of SRO with spatially varying shear strain by artificially imposing Sr shearing in the sinusoid form, which is constructed for examining the existence of flexoelectricity in SRO. **(c)** The plot of shear strain and associated strain gradient in the SRO supercell. **(d)** The off-center displacements of Ru and O, and the unit-cell electric dipoles extracted from the SRO supercell.

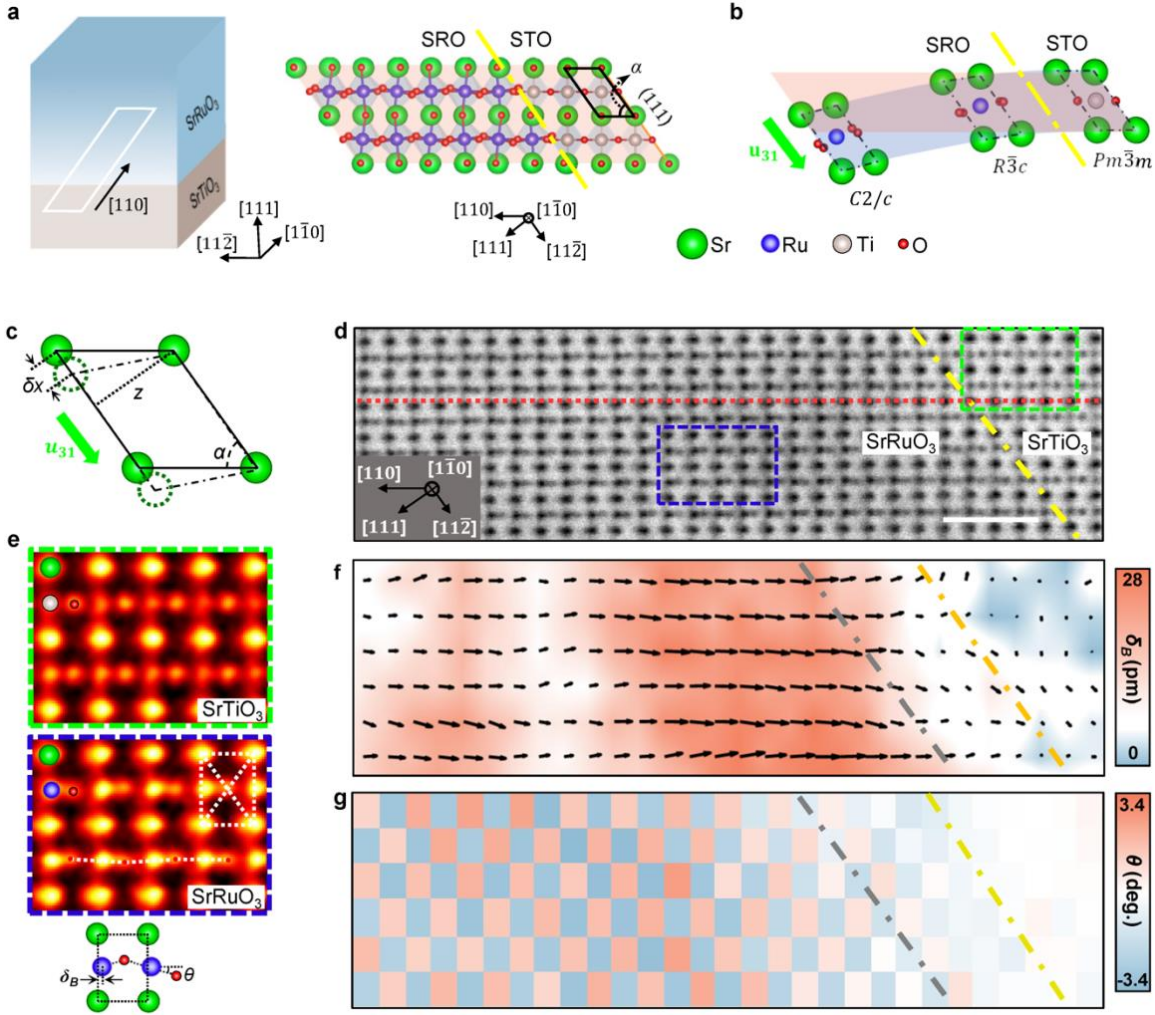

**Figure 2**. **Atomic-scale imaging of the shear strain gradient and flexoelectricity in SRO/STO (111). (a)** Schematic of the SRO/STO (111) heterostructure and the corresponding atomic structure coherently aligned along the [110] direction. **(b)** Schematic to show that a shear strain is induced in the heterostructure due to the low-symmetry structure distortion of the monoclinic SRO ($C2/c$) compared to the cubic STO substrate (($Pm\bar{3}m$) with a bridging rhombohedral phase of SRO ($R\bar{3}c$) at the interface. **(c)** Schematic to show the geometry for calculating the shear strain. **(d)** Cross-sectional ABF-STEM image of the SRO/STO (111) heterostructure, where the red dotted line is drawn along the STO [110] direction for reference to visualize Sr shear in SRO, and the yellow dash-dotted line marks the interface of SRO/STO. The scale bar is 1 nm. **(e)** Zoomed-in images of representative STO and SRO regions as marked by the green and blue dashed squares in (d), respectively. The bottom schematic shows the B-site cation displacement $\delta_B$ and the projected octahedral tilting angle $\theta$. **(f and g)** Mapping of B-site cation off-center displacements $\delta_B$ in reference to the Sr cations (f) and oxygen octahedral tilting (g) in (d).

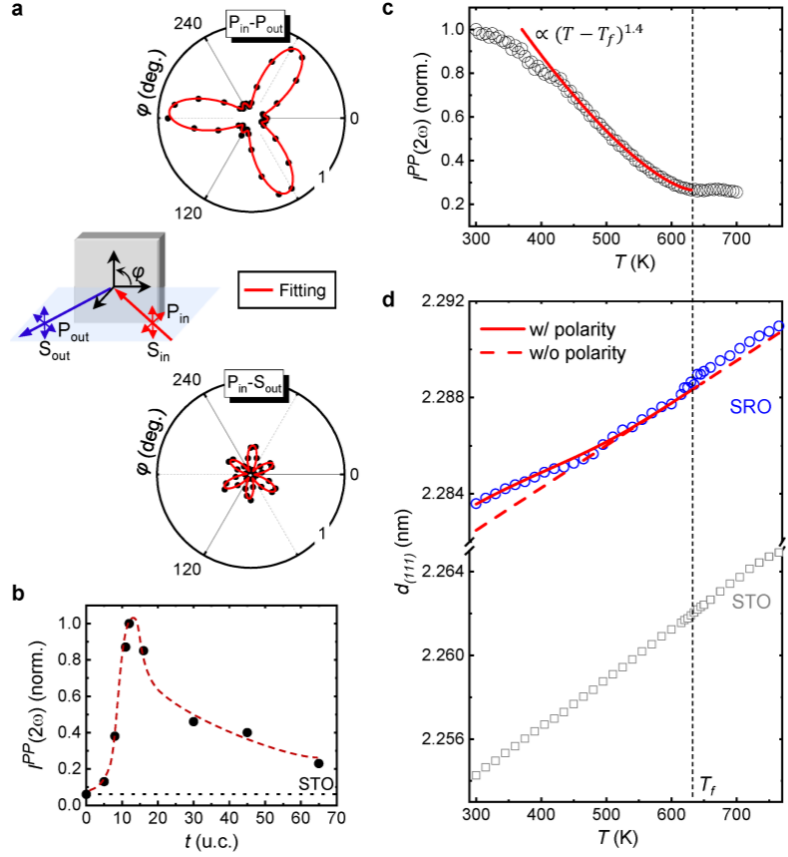

**Figure 3. Examination of the flexo-polar structure and transition of SRO by SHG and XRD.**
(a) SHG anisotropic patterns of an 11 u.c. SRO/STO (111) heterostructure acquired in $P_{in}$-$P_{out}$ (upper) and $P_{in}$-$S_{out}$ (lower) configurations. The schematic illustrates the configurations of SHG measurements. The red solid lines in the plots are the fitting curves based on the model of three degenerated polar monoclinic domains. (b) Thickness-dependent SHG intensity $I^{PP}(2\omega)$ of the SRO film. The red dashed line is drawn to guide the eye. The black dashed line marks the baseline signal of a STO substrate. (c) Temperature-dependent SHG intensity (black empty circles) suggests the polar to nonpolar phase transition, which can be described by a critical behavior (the red solid line). (d) Evolution of the (111) $d$-spacings of SRO (STO) measured by temperature-dependent XRD is plotted as blue (grey) empty circles (squares). The red solid and dashed lines show the predicted evolution of the temperature-dependent $d_{(111)}$ of SRO with and without polarity, respectively.

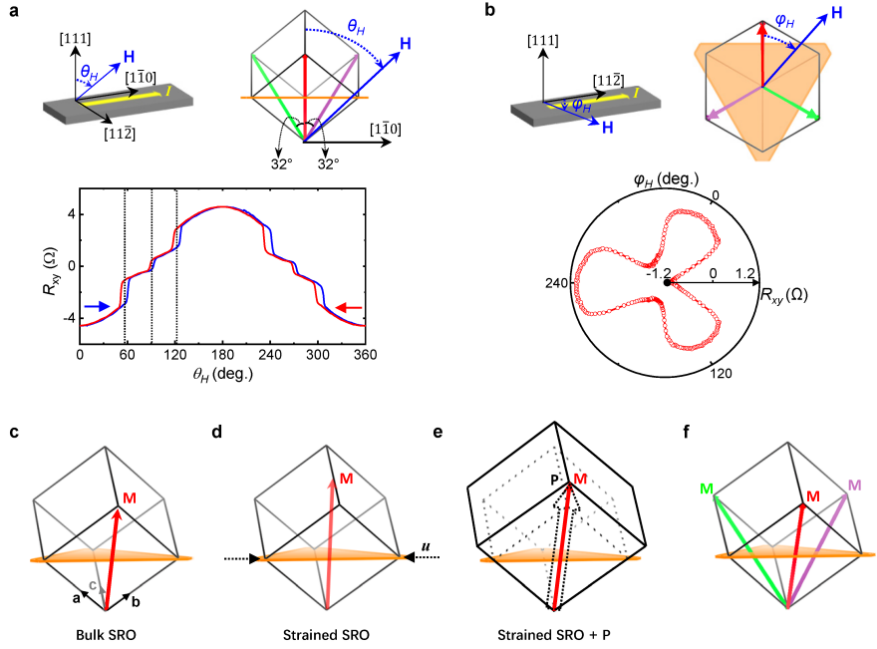

**Figure 4**. **Flexoelectric control of the magnetic anisotropy of SRO. (a** and **b)** Field angle-dependent Hall resistance measurement with the magnetic field $H$ rotating in the $(11\bar{2})$ plane (a) and in the (111) plane (b). $\theta_H$ and $\varphi_H$ refers to the field rotation angle in the two cases, respectively. **(c-e)** schematic to show the MEA (red arrow) of SRO with different phases: bulk *Pbnm* (c), strained *C2/c* without polarization (d), and *C2/c* with Ru off-centering induced electrostriction (e). **(f)** schematic to show the three MEAs (arrows in different colors) in the polar (111) SRO.

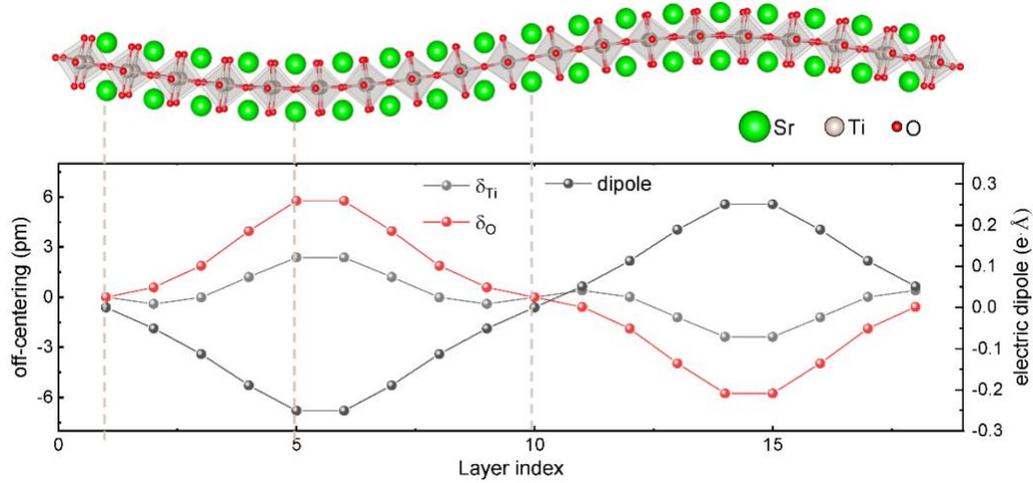

**Extended Data Figure 1**. **Theoretical simulation of flexoelectricity in STO under shear strain.** The supercell of STO was constructed in the same way as that used for SRO in the main article, thus possessing the same form of shear strain and associated gradient. The resulted ionic off-center displacements referencing to the Sr cage and corresponding unit-cell electric dipoles are extracted and plotted in the graph.

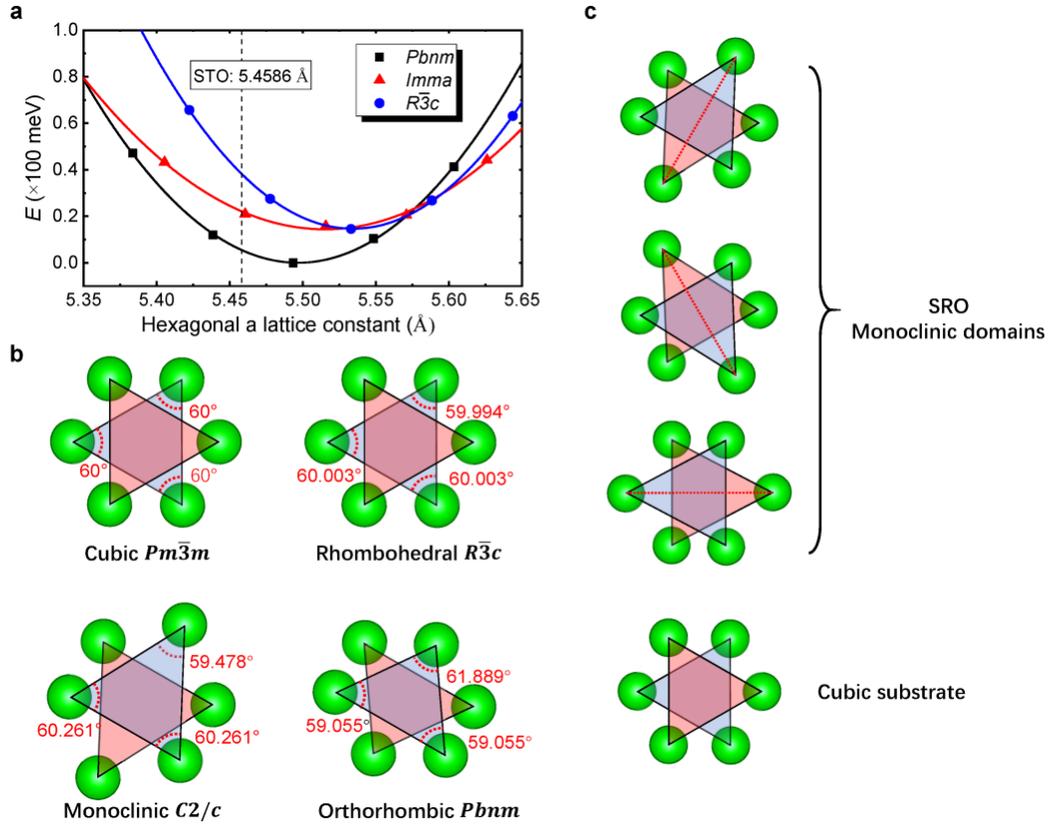

**Extended Data Figure 2**. **Calculated energy cost and the schematic honeycomb lattice of the different perovskite structures viewed from the [111] direction.** (**a**) Calculated energy cost of SRO in *Pbnm*, *Imma*, and $R\bar{3}c$ space groups under varying (111) epitaxial strains. Note that the *Imma* space group undergoes symmetry reduction to *C2/c* under compressive strain. (**b**) The cubic $Pm\bar{3}m$ structure of STO shows regular honeycomb shape while the rhombohedral $R\bar{3}c$, monoclinic $C2/c$, and orthorhombic $Pbnm$ structure of SRO show increasing distortion. (**c**) The low-symmetry SRO, e.g. *C2/c*, is expected to have three degenerated ferroelastic domains when epitaxially grown on a high-symmetry cubic substrate, which has three-fold rotational symmetry at the surface.

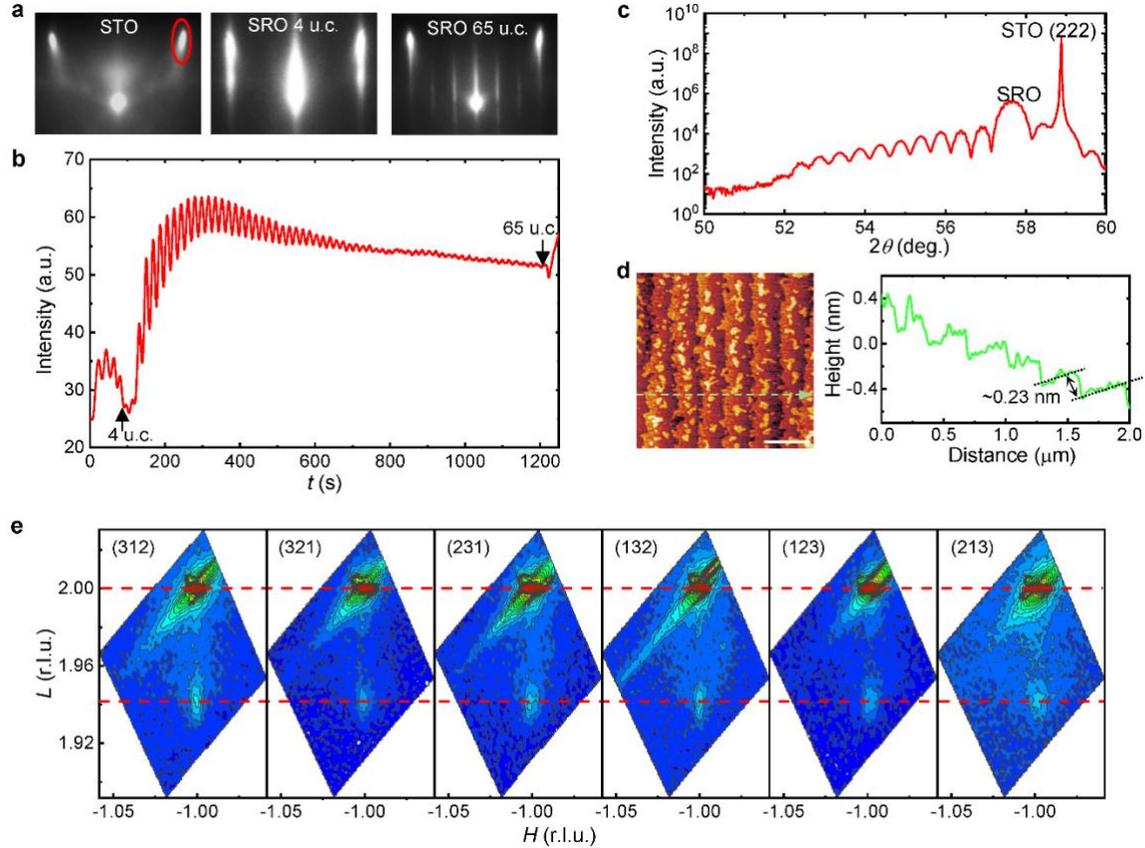

**Extended Data Figure 3**. **SRO film growth and characterization.** **(a)** In-situ RHEED patterns of the STO (111) substrate (a), 4 u.c. SRO film (b), and 65 u.c. SRO film (c) at 670 ℃ along the $[1\bar{1}0]$ azimuthal direction. **(b)** the RHEED intensity oscillation monitored from the diffraction spot, as marked by the red oval in (a), indicating a layer-by-layer growth mode of the SRO film. **(c)** $\omega$-$2\theta$ scan of the (222) plane of a 65 u.c. SRO/STO (111) heterostructure. The diffraction fringes indicate high quality of the heterostructure. **(d)** A representative AFM topography image of a 65u.c. SRO film with unit-cell step-terrace morphology inherited from the STO substrate demonstrates well-controlled growth. **(e)** RSM of the symmetric {123} planes of 65 u.c. SRO/STO (111). The same position of SRO {123} indicates a macroscopic 3-fold rotational symmetry of the polar SRO structure.

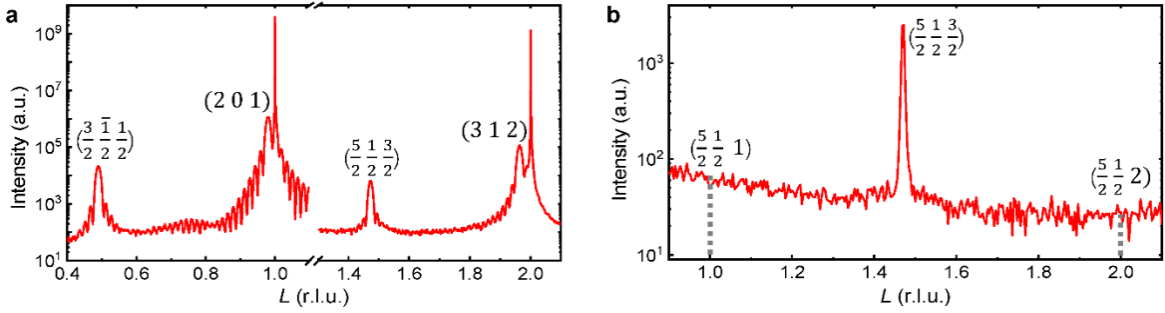

**Extended Data Figure 4**. **Half-integer Bragg diffraction of the (111) SRO film**. The ½(odd, odd, odd)-type diffraction peaks in **(a)** indicate the existence of out-of-phase tilting while the absence of ½(odd, odd, even)-type peaks in **(b)** suggest the suppression of the inherent in-phase tilting. Therefore, the results demonstrate a tilting pattern of $a^-a^-c^{0+\delta}$ ($\delta$ is either close to 0 or negative) in the (111) SRO film.

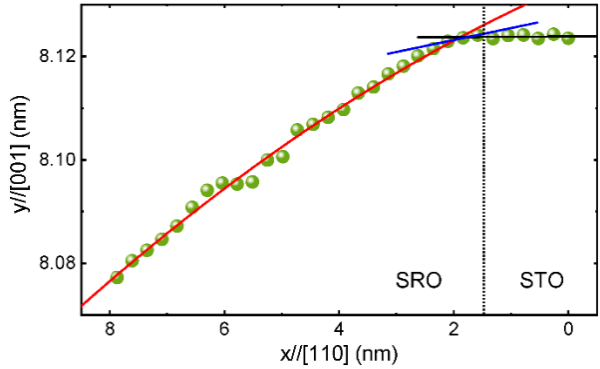

**Extended Data Figure 5**. **Analysis of shear strain in STEM.** The green dots represent the extracted Sr position in the SRO/STO (111) heterostructure. The red line shows a polynomial fitting with the function $y = 8.13377 - 0.00477x - 2.98002 \times 10^{-4}x^2$. The fitting describes well the trend of Sr shearing in the bulk of SRO film, but deviates in the interfacial region. The blue line shows the linear fitting of the interfacial region, giving a slope of 0.0023. The black line shows the linear fitting of the STO region, giving a slope of 0.00005. Accordingly, the change of the shearing angle, $\delta\alpha$, in reference to STO can be extracted from the derivative. The bulk strain gradient can be then estimated from the slope of the polynomial function, yielding a value of $\sim 2.0 \times 10^6$ m$^{-1}$.

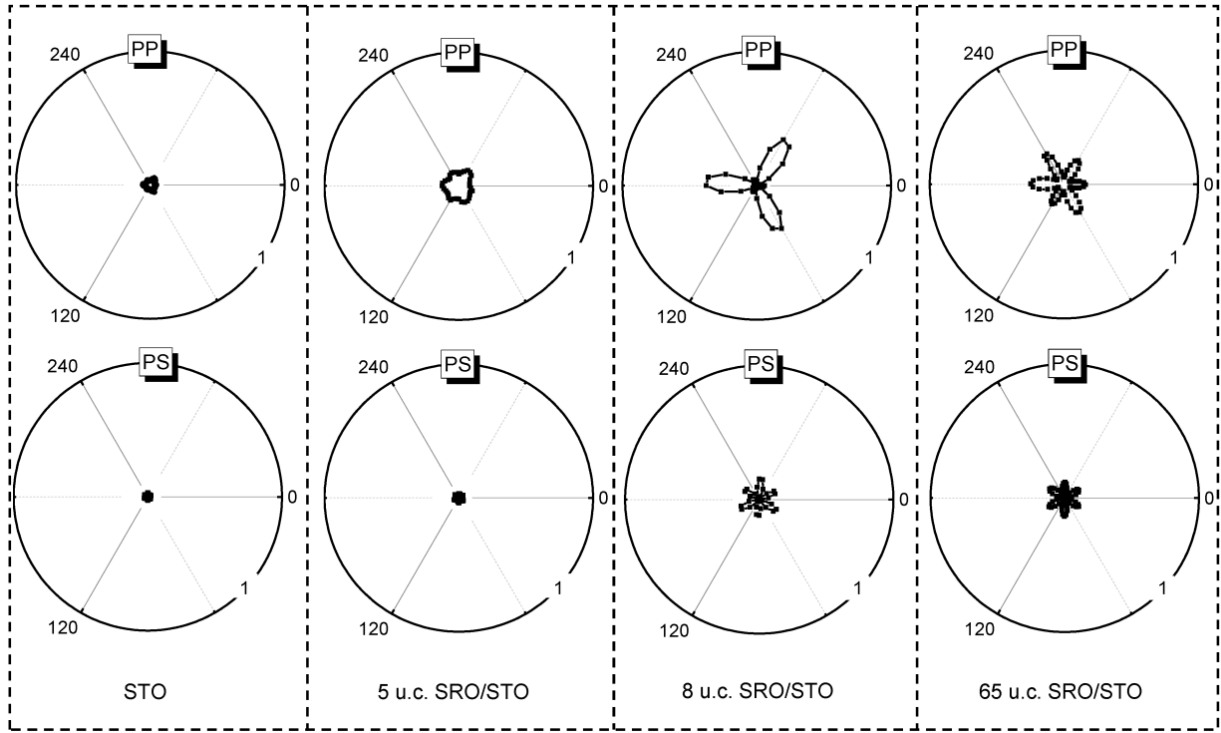

**Extended Data Figure 6. SHG of the SRO/STO (111) with different SRO thicknesses in the $P_{in}$-$P_{out}$ (upper) and $P_{in}$-$S_{out}$ (lower) configurations.** The three-fold symmetric SHG of STO with a very weak intensity should come from the surface, which naturally breaks inversion symmetry. The SHG of 5 u.c. SRO/STO has both a distorted pattern and weak intensity, indicating the polarity of 5 u.c. SRO is strongly suppressed.

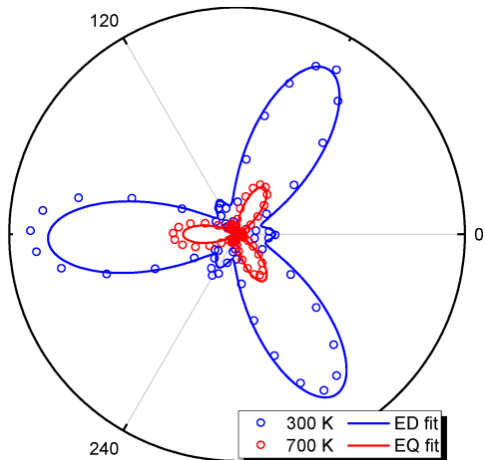

**Extended Data Figure 7. Temperature-dependent SHG patterns of SRO/STO (111).** The SHG pattern (red empty circles) at 700 K is fitted with electric quadrupole (EQ) contribution of point group of $\bar{3}m$ (the red solid line). The pattern at 300 K (blue empty circles) is fitted with the electric dipole (ED) contribution of point group $m$ with three equivalent domains (blue solid line).

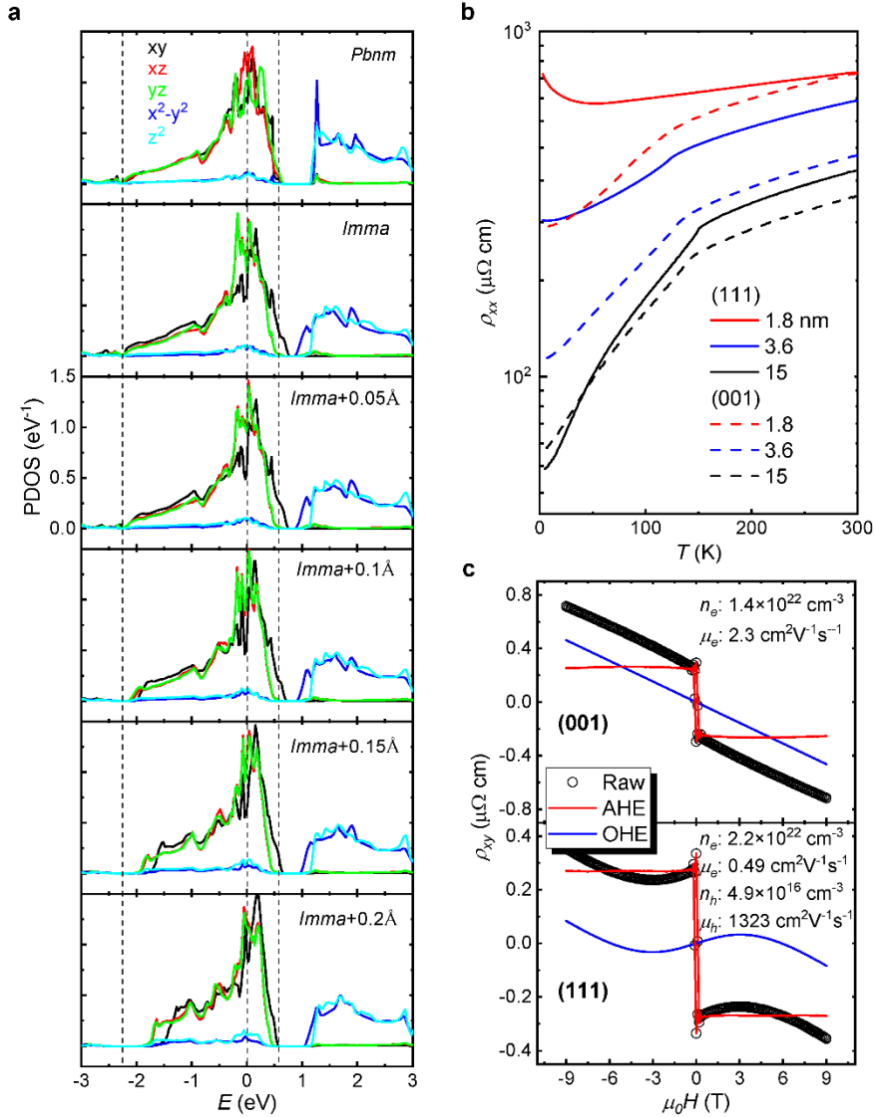

**Extended Data Figure 8. PDOS and transport study of polar and non-polar SRO films with different thicknesses**. (**a**) Calculated PDOS of SRO in different structures: *Pbnm*, *Imma*, *Imma* with Ru off-centering displacements. Here we show the calculated results based on the relaxed *Imma* phase without epitaxial strain, focusing on the effect of Ru displacements, in which we found that the strain does not change the overall trend. (**b**) Longitudinal resistivity of (111) and (001) SRO films with different thicknesses. All the (111) SRO films remain metallic. The 15 nm-thick (111) SRO film shows a large residual resistivity ration (RRR) of ~7.3, compared to ~5.3 of the (001) SRO film with the same thickness. An additional characteristic of the polar SRO film is its relatively higher resistivity compared to that of the non-polar (001) SRO film, which can be ascribed to two factors: first, the reduced electronic bandwidth due to Ru off-centering amplifies electron scattering; second, with film thickness decreasing, defects begin to take a dominant role, particularly for the [111] orientation due to its charged layer configuration. (**c**) Transverse resistivity of SRO films with a thickness of 3.6 nm along both orientations at 90 K. The high-field ordinary Hall component was fitted with either a single-carrier or two-carrier transport model. The fitting parameters are given in the plot.

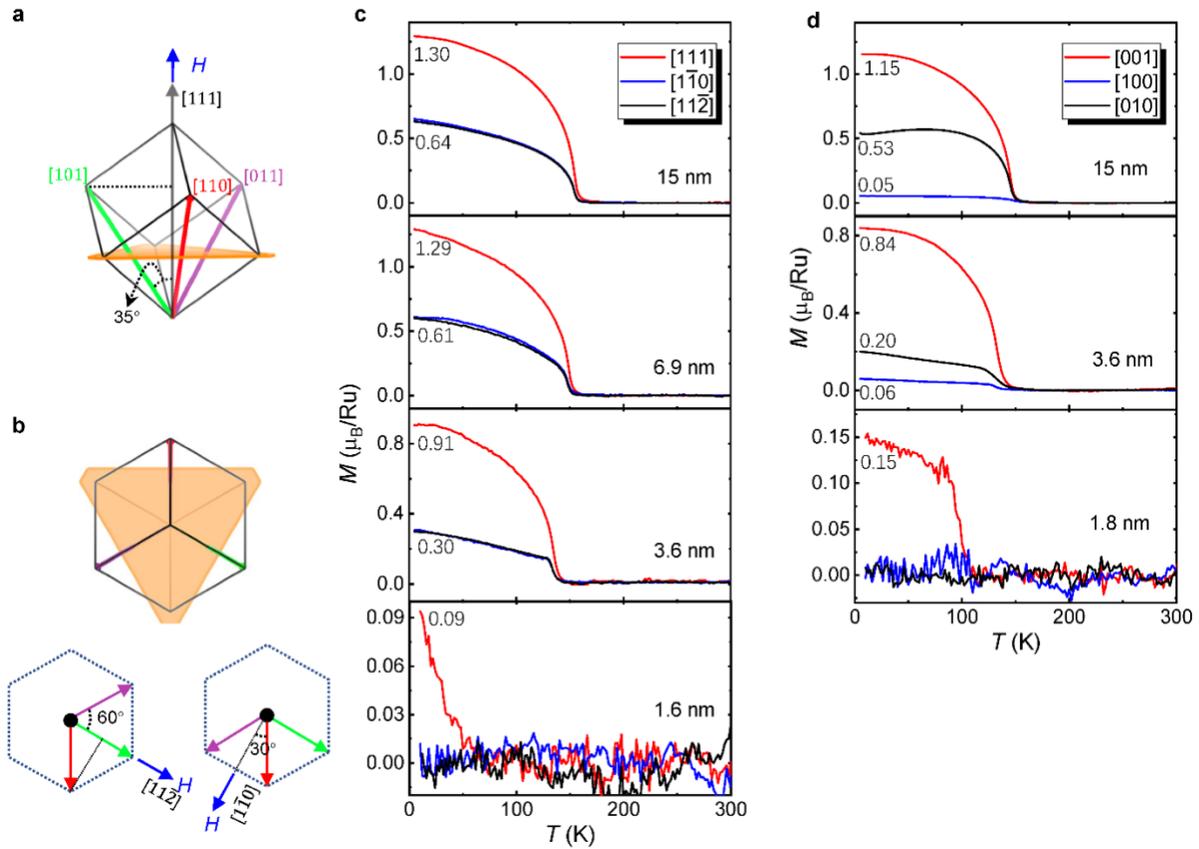

**Extended Data Figure 9**. **Magnetization measurement of the polar (111) and the non-polar (001) SRO films**. (**a** and **b**) Schematic of the out-of-plane (a) and in-plane (b) magnetization measurement for the (111) SRO. (**c** and **d**) IP and OOP magnetization-temperature (MT) curves of the (111) SRO films (c) and the (001) SRO films (d) with varying thicknesses. The saturated magnetic moments are marked in the figure. Therefore, it can be seen that the maximum $M_{IP}/M_{OOP}$ ratio of (111) SRO is ~0.49, 0.47, 0.34 for 15, 6.9, 3.6 thick films, respectively, and purely OOP in the 1.6 nm thick film, indicating an increasing OOP magnetic component in thinner films. Such a trend suggests a different structure with pure OOP magnetization at the SRO/STO (111) interface. This magnetic structure is supposed to originate from the interfacial rhombohedral SRO (~5 u.c.), which lacks flexoelectricity. In addition, it is worth noting that the saturated magnetic moment of the 15 nm-thick (111) SRO film along the magnetic easy axis can be derived to be as large as ~1.58 μ$_B$/Ru, by contrast to 1.27 μ$_B$/Ru of the (001) SRO film with the same thickness. The ferromagnetic $T_C$ of the (111) SRO film is ~157 K, also representing a sizable enhancement compared to 146 K of the (001) SRO film. The enhancement of ferromagnetism in (111) SRO is solid even compared with the highest-quality (001) films that were reported previously[45,46]. The enhancement gradually diminishes with film thickness decreasing, probably caused by the influential role of defects in thinner films.